\documentclass{iopart}

\begin{document}
\eqnobysec
\title{One-loop $f(R)$ Gravitational Modified Models}

\author{Guido Cognola  and Sergio  Zerbini}
\address{Dipartimento di Fisica, Universit\`a di Trento and Gruppo
 Collegato INFN \\ 
Trento, 38050 Povo, Italia}
\ead{\mailto{cognola@science.unitn.it}, \mailto{zerbini@science.unitn.it}}

\begin{abstract}
The one-loop quantisation of a general class of modified gravity 
models around a classical de Sitter background is presented.
Application to the stability of the models is addressed.

\end{abstract}

\section{Introduction}

In contrast with what expected by the so called standard cosmological model,
quite surprisingly recent astrophysical data indicate that the 
Universe is currently in a phase of accelerated expansion.
This has been seen by the observation of the
light curves of several hundred of type $I^a$ supernovae \cite{SN} and also
emerges by a detailed analysis of the
Cosmic Microwave Background as recently measured by WMAP satellite 
and other experiments \cite{CMB}.

The associated  theoretical issue, called the Dark Energy problem, 
might be solved in different ways. 
The simplest one consists in the introduction of suitable 
new cosmological ``matter'' fields with unusual equation of state
(quintessence, phantom, k-essence). 
Another possibility consists in the modification of General Relativity 
by adding new gravitational higher order derivative and non linear terms 
in the curvature invariants to the Einstein-Hilbert action.
Finally, the accelerated expansion of the Universe can also be accounted for
within the framework of non-perturbative renormalisation of
quantum gravity \cite{reuter}.   

In this paper, we will be mainly interested in this second possibility. 
A short motivation may be presented as follows:  the inclusion of terms 
which grow up when curvature decreases, 
for example inverse curvature terms \cite{turner}, 
which may origin from string/M-theory \cite{nojiri0}, may explain 
such current accelerated expansion and may give negligible 
contributions to early cosmology. 
However, such modified gravity models with Einstein-Hilbert term plus
inverse curvature terms contain some instabilities \cite{ins} and do not 
pass solar system tests, but their further modification by higher 
derivative curvature squared terms could improve their behaviour \cite{NO2}.
In this respect, it has to be note that in general such kind of models
are not renormalisable and for this reason they have to be seen not as
fundamental, but as effective theories. In such cases the models
can be made renormalisable, but the prise to pay is the lack of unitarity
\cite{stelle,barth}. 
Furthermore, if Einstein's gravity is only an effective theory, 
then at the early Universe the (effective) quantum gravity 
should be different from Einstein's one.

The widely discussed possibility in this direction is quantum $R^2$ gravity
(for a review, see \cite{buch}).
However, other modifications deserve attention,
because they may produce extra terms, which may help to realise the early time 
inflation. This is supported by the possibility of accelerated expansion
with simple modified gravity. 

Here, we will report the general $f(R)$ gravity 
at one-loop level in de Sitter Universe \cite{f}.
Such a program for the case of Einstein's gravity has been initiated 
in \cite{perry,duff80,frad} (see also \cite{ds1}). 
Also in that case the theory is multiplicatively non-renormalisable.

Using generalised zeta-function regularisation 
(see, for example \cite{eli94,byts96}), 
it is possible to get the one-loop effective action 
and to study the possibility of stabilisation of de Sitter 
background by quantum effects. Moreover, such an approach 
may suggest the way how to resolve the cosmological constant 
problem \cite{frad}. 
Hence, the study of one-loop $f(R)$ gravity is a quite natural
step in the realisation of such a program, having in mind, that consistent 
treatment of quantum gravity does not exist yet.

\section{Models for Dark Energy}

As mentioned in the Introduction, it is possible to accommodate the current 
cosmic acceleration without modify
the Einstein-Hilbert action, but in such a case one has to introduce by hands 
a cosmic fluid with negative pressure, 
realizing in this way a naive model for the Dark Energy component.

It is also well known that the simplest theoretical possibility 
of this kind consists in the introduction, 
in the standard cosmological model, of a
{\em positive cosmological constant}. It is a historical  fact that 
the cosmological constant was introduced
for the first time in GR by Einstein himself, in order to get 
a static cosmological model and immediately abandoned 
after the observation of cosmic expansion. 
Now it appears again but, in some sense, for the opposite reason.

As it is well known, within the framework of quantum field theory
in curved space-time, a constant cosmological term can be identified 
with the vacuum energy, but unfortunately, the estimation of such a 
quantity gives rise to a value extremely large if
compared with the one expected for the cosmological constant. 
In fact, within the quantum field theory, the vacuum energy is divergent. 
A cutoff at the Planck or electroweak scale leads 
to a cosmological constant which is, respectively, 
$10^{123}$ or $10^{55}$ times larger than the observed value, 
$\Lambda/8\pi G \simeq 10^{-47}$ GeV$^4$.
This is the well known {\em cosmological constant problem}, which probably
might be solved in a unified theory of all interactions. 

Other possibility consists in using a cosmological scalar field, 
a sort of ``dynamical cosmological constant'', as for example
{\em quintessence}, if the rate pressure/density is greater than $-1$
or {\em phantom} if such a rate is less than $-1$.
We remind that for the cosmological constant the rate 
pressure/density is exactly $-1$.

\section{The $f(R)=R-\mu^4/R$ model}

Now let us discuss the second possibility, i.e. the fact that cosmic 
acceleration can also be explained  by adding other non linear terms 
to the Einstein-Hilbert Lagrangian.
For example, it is well known  that quadratic terms in the curvature, 
may be induced by quantum effects associated with conformally 
coupled matter fields and these effects modify the 
cosmological solutions of Einstein's equations at {\em early time}. 
In this way one could obtain, 
for example, a de Sitter inflationary phase \cite{staro}.

On the contrary, by considering terms depending on the 
inverse of the curvature, one modifies the solutions 
at {\em present time} \cite{turner}.  
In this case, one has to pay attention and try to modify  
GR in such a way that 
astrophysical tests are not violated, since GR is in excellent agreement 
with astrophysical data and one has to save also early cosmology, which is
in good agreement with observations.

We illustrate this second possibility by considering the simplest model 
introduced in  \cite{turner} and defined by the action 
\begin{eqnarray}
S=\frac{1}{16\pi G}\int\,d^4x\,\sqrt{-g}\,\left( R-\frac{\mu^4}{R}\right)\,
+S_{matter}\,.\nonumber
\end{eqnarray}
This is the action in the so called {\em matter frame} and this gives rise to
$4^{th}$ order field equations.

By making use of a suitable conformal transformation, one may pass to the so 
called {\em Einstein frame}, in which  
the action assumes the Einstein-Hilbert form and the gravitational  
additional degrees of freedom are represented 
by a scalar field $\phi$ with a  complicated potential, which reads 
\begin{eqnarray} 
V(\phi)=\frac{\mu^2}{8\pi G}\,\,
\left( e^{-\phi/\sqrt{12\pi G}}-1\right)^{1/2}\,\,
e^{-\phi/\sqrt{3\pi G}}\,.
\nonumber\end{eqnarray} 
Depending on the initial values of $\phi$ one can have different
solutions.
For a special value of $\phi(0)$, one has a de Sitter solution, but
this is unstable and requires fine tuning in order have 
corrections to standard cosmology starting 
at the present epoch. Moreover, it is in contrast with
gravitational tests on solar system.
For this reason, different models have been considered.

\section{Arbitrary $f(R)$ modified gravitational models}

The starting point is the classical action depending on a generic function 
of the scalar curvature $R$
\begin{eqnarray}
S=\frac{1}{16\pi G}\int\,d^4x\,\sqrt{-g}\,f(R)\,,\nonumber
\end{eqnarray}
which gives rise to the field equations
\begin{eqnarray}
f'(R)R_{\mu\nu}-\frac{1}{2} f(R)g_{\mu\nu}+\left(
\nabla_\mu \nabla_\nu-g_{\mu\nu} \Delta\,\right) f'(R)=0\,.\nonumber
\end{eqnarray}
Requiring the existence of solutions with constant scalar curvature 
$R={\hat R}$,
one obtains the conditions \cite{barrow}
\begin{eqnarray}
2f({\hat R})={\hat R}\,f'({\hat R})\,, 
\qquad\qquad
R_{\mu\nu}=\frac{f({\hat R})}{2 f'({\hat R})}g_{\mu\nu}=
\frac{{\hat R}}{4}g_{\mu\nu}\,. \nonumber
\end{eqnarray}
and this means that these solutions are Einstein's spaces with an 
effective cosmological constant 
\begin{eqnarray}
\Lambda_{eff}=\frac{f({\hat R})}{2f'({\hat R})}=\frac{{\hat R}}{4}\,.\nonumber
\end{eqnarray}
Such a class of constant curvature solutions contains 
black holes in the presence of a non vanishing cosmological
constant\cite{vanzo2}, like the Schwarzschild-(anti)de Sitter
and all the topological solutions associated with a
negative $\Lambda_{eff}$\cite{Vanzo}. 
In general, their black hole entropies 
do not satisfy the area law \cite{vanzo2}. 


\section{Quantum field fluctuations around a maximally symmetric space}

In this Section, we will present a summary of the one-loop 
quantisation of the general modified models 
in the Euclidean signature. We shall make use of the 
background field method and 
zeta-function regularisation, having in  mind that, in general,  
these models are not renormalisable and one is dealing only 
with an effective approach. 

To begin with, we recall that  the condition 
\begin{eqnarray}2\hat f=\hat R \hat f'\,,\qquad\qquad
\hat f=f(\hat R)\,,\qquad\hat f'=f'({\hat R})\,,\nonumber
\end{eqnarray}
ensures the existence of constant curvature solutions and in particular
maximally symmetric spaces solutions like for, example, the
de Sitter space,  the one we are interested in.

For maximally symmetric spaces, the Riemann and Ricci tensors are given by 
the expressions
\begin{eqnarray}
\hat R_{ijrs}=\frac{{\hat R}}{12}\left( \hat g_{ir}\hat g_{js}-
\hat g_{is}\hat g_{jr}\right) \:,\qquad
\hat R_{ij}=\frac{{\hat R}}{4}\,\hat g_{ij}\,,\qquad
R={\hat R}\,.\nonumber
\end{eqnarray}
Now we expand the metric and all quantities in the action around the
maximally symmetric solution, that is
\begin{eqnarray} g_{ij}=\hat g_{ij}+h_{ij}\:,\qquad
g^{ij}=\hat g^{ij}-h^{ij}+h^{ik}h^j_k+...\qquad
h=\hat g^{ij}h_{ij}\:,\nonumber 
\end{eqnarray}
and up to second order in $h_{ij}$
\begin{eqnarray}
\frac{\sqrt{g}}{\sqrt{\hat g}}=1+\frac12h+\frac18h^2-\frac14h_{ij}h^{ij}
+{\cal O}(h^3)\nonumber  
\end{eqnarray}
\begin{eqnarray}
 R &\sim& {\hat R}-\frac{{\hat R}}{4}\,h+\nabla_i\nabla_jh^{ij}-\Delta\, h
+\frac{{\hat R}}{4}\,h^{jk}h_{jk} 
\nonumber \\ && 
-\frac14\,\nabla_ih\nabla^ih
-\frac14\,\nabla_kh_{ij} \nabla^kh^{ij} +\nabla_ih^i_k\nabla_jh^{jk}
-\frac12\,\nabla_jh_{ik}\nabla^ih^{jk} \:. \nonumber
\end{eqnarray}
Here the symmetric tensor $h_{ij}$ has to be considered as a small
fluctuation about the background metric $\hat g_{ij}$.

For technical reasons, it is convenient to carry 
out the standard expansion of the tensor
field $h_{ij}$ in irreducible components, namely
\begin{eqnarray}
h_{ij}=\hat h_{ij}+\nabla_i\xi_j+\nabla_j\xi_i+\nabla_i\nabla_j\sigma
+\frac14\,g_{ij}(h-\Delta\,\sigma)\:,
\nonumber\end{eqnarray}
where $\sigma$ is the scalar component, 
while $\xi_i$ and $\hat h_{ij}$ are the vector
and tensor components with the properties
\begin{eqnarray}
\nabla_i\xi^i=0\:,\qquad\qquad \nabla_i\hat h^i_j=0\:,
\qquad\qquad\hat h^i_i=0\:. 
\nonumber\end{eqnarray}
In terms of the irreducible components
of the $h_{ij}$ field, the quadratic part of the Lagrangian density, 
disregarding total derivatives, becomes
\begin{eqnarray}
{\cal L}_2&=&
\frac{1}{12}\,\hat h^{ij}\,(3\hat f\Delta\,-3\hat f+\hat R\hat f'\,)\,\hat h_{ij}
\nonumber\\&&\qquad
+\frac1{16}\,(2\hat f-\hat R\hat f')\:\xi^i\,(4\Delta\,+\hat R)\,\xi_i
\nonumber\\&&
+\frac{1}{32}\:h\left[ 9\hat f''\,\Delta\,^2
   -3(\hat f'-2R\hat f''\,)\,\Delta\,
     +2\hat f-2\hat R\hat f'+\hat R^2\hat f''\,
\right]\,h
\nonumber\\&&
+\frac{1}{32}\:\sigma\,\left[ 9\hat f''\Delta\,^4 
-3(\hat f'-2\hat R\hat f''\,)\,\Delta\,^3\right.
\nonumber\\&&\qquad\qquad\left.
     -(6\hat f-2\hat R\hat f'-\hat R^2\hat f''\,)\,\Delta\,^2
      -\hat R(2\hat f-\hat R\hat f'\,)\,\Delta\,
\right]\,\sigma
\nonumber\\&&
+\frac{1}{16}\:h\left[ -9\hat f''\,\Delta\,^3
     +3(\hat f'-2\hat R\hat f''\,)\,\Delta\,^2
      +\hat R(\hat f'-\hat R\hat f''\,)\,\Delta\,
\right]\,\sigma\:.
\nonumber\end{eqnarray}
In order to quantise the model, now one has to add {\em gauge fixing} 
and {\em ghost contributions}. 
Such terms are quite complicated, 
but with the help of a tensor manipulations program, 
we were able to obtain the one-loop effective action 
(here written in the Landau gauge)
(on-shell condition: $X=(2\hat f-\hat R\hat f')/4=0$)
\begin{eqnarray}
\Gamma_{off-shell}&=&\frac{24\pi\hat f}{G\hat R^2}
+\frac12\log\det\left(-\Delta\,_2-\frac{\hat R}{6}\,\,\frac{X+2\hat f}{X-2\hat f}\right)
\nonumber \\&&
       -\frac12\log\det\left(-\Delta\,_1-\frac{\hat R}{4}\right)
        -\frac12\log\det\left(-\Delta\,_0-\frac{\hat R}{2}\right)
\nonumber\\ &&
   +\frac12\log\det\left\{\left(-\Delta\,_0
    -\frac{5\hat R}{12}-\frac{X-2\hat f}{6\hat R \hat f''}\right)^2\,
\right.\nonumber\\ &&\left.\qquad
   -\left[\left(\frac{5\hat R }{12}+\frac{X-2\hat f}{6\hat R\hat f''}\right)^2
-\frac{\hat R^2}{6}-\frac{X-\hat f}{3\hat f''} \,\right]\right\}\,,
\nonumber\end{eqnarray}

\begin{eqnarray} 
\Gamma_{on-shell}&=&\frac{24\pi\hat f}{G\hat R^2}
+ \frac12\,\log\det\ \left[ \ell^2
\left(-\Delta\,_2+\frac{R_0}6\right)\right] \nonumber\\&&
      -\frac12\,\log\det\left[ \ell^2 \left(-\Delta\,_1-\frac{\hat R}4\right) \right]
\nonumber\\&&
    +\frac12\,
\log\det \left[ \ell^2 \left(
-\Delta\,_0-\frac{\hat R}3+\frac{2\hat f}{3\hat R\hat f''}\right)\right]\,. 
\label{GAon}
\end{eqnarray}

On the de Sitter manifold (more precisely $SO(4)$, 
since we work in the Euclidean section) 
the eigenvalues of the Laplace operator are known
and this means that zeta-functions are exactly computable
and so we can obtain the
one-loop effective action in the closed form
\begin{eqnarray}
\Gamma=\Gamma({\hat R})=
\frac{24\pi}{G\hat R^2}\,f(\hat R)+
F_1(\hat R)+F_2(\hat R)\,\log\frac{\ell^2\hat R}{12}\,.
\nonumber\end{eqnarray}
Here $F_1(\hat R)\,,\,F_2(\hat R)$ are complicated, but known functions 
of the scalar curvature $\hat R$. 

\section{Stability of de Sitter solution in $f(R)$ models}

The one-loop effective action may be used to investigate the role of quantum 
corrections of these modified gravitational models 
to the background cosmology (see, for example, \cite{f}). 
In the following, we would like to present an 
application to the stability of such models.
 
In fact, the stability of the de Sitter solution may be obtained
by imposing the one-loop effective action to be real and this happens 
if the differential Laplace-like operators, which determine the 
effective action,  do no possess negative eigenvalues. 
The eigenvalues of Laplace-like  operators in $SO(4)$ of the kind
$L_i=-\Delta\,_i+c_i \hat R$, can be evaluated,  recalling that 
for the pure Laplacians one has
\begin{itemize}
\item $-\Delta\,_0$: $\qquad (n^2+3n)(\hat R/12)\,,\qquad n=0,1,2,...$
\item $-\Delta\,_1$: $\qquad (n^2+5n+3)(\hat R/12)\,,\qquad n=0,1,2,...$
\item $-\Delta\,_2$: $\qquad (n^2+7n+8)(\hat R/12)\,,\qquad n=0,1,2,...$
\end{itemize}
Then, from equation (\ref{GAon}), we get the following 
conditions, which state the stability of de Sitter solution:
\begin{eqnarray} 
&&2f(\hat R)-\hat R\,f'(\hat R)=0\,,\qquad\qquad
\frac{f(\hat R)}{2\,f'(\hat R)}>0\,,\nonumber \\
&&\frac{2f(\hat R)}{\hat R^2\,f''(\hat R)}
=\frac{f'(\hat R)}{\hat R\,f''(\hat R)}>1\,.
\nonumber\end{eqnarray}
The first two equations ensure the existence of a solution with
positive constant curvature, while the third one ensures the stability
of such a solution. Such a condition has been obtained 
in \cite{faraoni} by a classical perturbation method.  

For example, for the model
\begin{eqnarray}
f(R)=R-\frac{\mu^4}{R}\,,
\nonumber\end{eqnarray} 
one has
\begin{eqnarray}
f'(R)=1+\frac{\mu^4}{R^2}\,,\qquad f''(R)=-2\frac{\mu^4}{R^3}\,, 
\nonumber\end{eqnarray} 
and $\hat R=\sqrt{3}\mu^2$. As a result, the model is always unstable.

For the slightly modified model \cite{dick}
\begin{eqnarray}
f(R)=R-\frac{\mu^4}{R}+aR^2 \,,
\nonumber\end{eqnarray} 
one has again $\hat R=\sqrt{3}\mu^2$, but now
\begin{eqnarray}
f'(R)=1+\frac{\mu^4}{R^2}+2aR\,,\qquad f''(R)=-2\frac{\mu^4}{R^3}+2a\,. 
\nonumber\end{eqnarray} 
A direct calculation 
leads to the stability condition
$a>1/3\sqrt{3}\mu^2$, in agreement with \cite{nojiri0,faraoni}.

Finally in the case
\begin{eqnarray}
f(R)=R+aR^2-2\Lambda\,,
\nonumber\end{eqnarray}
one has $\hat R=4\Lambda$, and
\begin{eqnarray}
f'(R)=1+2aR\,,\qquad f''(R)=2a\,. 
\nonumber\end{eqnarray} 
As a result, the model is stable for $a>0$ \cite{barrow}.

\section{Conclusions}

Generalising a previous program concerning the one-loop Einstein's gravity 
in the de Sitter background \cite{frad},
we have here presented the one-loop effective action for
a general $f(R)$ gravitational modified model. 
This one-loop effective action may be used to investigate 
the role of quantum corrections in cosmology.

Furthermore, as a non trivial application,  we have derived 
the condition which ensures the stability of 
the de Sitter solution in such a class of modified gravity theories. 
Such a condition is in full agreement with
the one obtained in \cite{faraoni}, 
where the covariant and gauge-invariant 
formalism of Bardeen-Ellis-Bruni-Hwang \cite{B} has been used.

We have also seen that generalising the simplest model in \cite{turner},
it is possible to build up models with a stable de Sitter solution,
which in principle could explain the recent cosmological data. 
From this point of view, models depending on a suitable function
of the Gauss-Bonnet invariant seem more promising, since they 
pass solar system tests for any reasonable choice of the 
function \cite{odi2005}.


\section*{References}

\begin{thebibliography}{99}

\bibitem{SN} A.G. Riess {\em et al.}, {\em Astron. Astrophys.} 
{\bf 116}, 1009 (1998); {\em Astron. J.} {\bf 118}, 2668 
(1999);
{\em Astrophys. J.} {\bf 560}, 49 (2001);
{\em Astrophys. J.} {\bf 607}, 665 (2004);
S. Perlmutter {\em et al.}, {\em Nature} {\bf 391}, 51 (1998);
{\em Astrophys. J.} {\bf 517}, 565 (1999); 
J.L. Tonry {\em et al.}, {\em Astrophys. J.} {\bf 594}, 
1 (2003);
R.  Knop {\em et al.}, {\em Astrophys. J.} {\bf 598}, 102 
(2003);
B. Barris {\em et al.}, {\em Astrophys. J.} {\bf 602}, 571 (2004).

\bibitem{CMB} 
A.D. Miller {\em et al.}, {\em Astrophys. J. Lett.} {\bf 524}, 
 L1 (1999); 
P. de Bernardis {\em et al.}, {\em Nature} {\bf 400}, 955 (2000);
A.E. Lange {\em et al.}, {\em Phys. Rev. D} {\bf 63}, 042001 (2001);
A. Melchiorri, L. Mersini, C.J. Odman and M. Trodden, {\em 
Astrophys. J. Lett.} {\bf 536},   
L63 (2000); 
S. Hanany {\em et al.}, {\em Astrophys. J. Lett.} {\bf 545},  
L5 (2000); 
D.N. Spergel {\em et al.}, {\em Astrophys. J. (Suppl.)} {\bf 
148},  175 (2003);
C.L. Bennett {\em et al.}, {\em Astrophys. J. (Suppl.)} {\bf 
148}, 1 (2003);
T.J. Pearson {\em et al.}, {\em Astrophys. J. } 
{\bf 591}, 556 (2003);
A. Benoit {\em et al.},  {\em Astron. Astrophys.} 
{\bf 399}, L25  (2003). 


\bibitem{reuter}
  A.~Bonanno and M.~Reuter,
  Int.\ J.\ Mod.\ Phys.\ D {\bf 13} (2004) 107.
  A.~Bonanno, G.~Esposito, G.~Rubano and P.~Scudellaro,
  ``The accelerated expansion of the universe as a crossover phenomenon,''
  arXiv:astro-ph/0507670.



\bibitem{turner} S. Capozziello, S. Carloni and A. Troisi,
Recent Research Developments in Astronomy and  Astrophysics-RSP/AA/21-2003,
 astro-ph/0303041;
S. M. Carroll, V. Duvvuri, M. Trodden and M. Turner, 
Phys.~Rev. D {\bf 70}, astro-ph/0306438.



\bibitem{nojiri0}
S. Nojiri, S. D. Odintsov, 
{\sl  Phys. Letters} {\bf B 576} (2003) 5, hep-th/0307071;

\bibitem{ins} T. Chiba, Phys.Lett. B575 (2003) 1, astro-ph/0307338;
A.D. Dolgov and M. Kawasaki, Phys.Lett. B573 (2003) 1, astro-ph/0307285;
M.E. Soussa and R.P. Woodard, Gen.Rel.Grav. 36 (2004) 855, astro-ph/0308114.


\bibitem{NO2} S. Nojiri and S.D. Odintsov,{\sl Phys.Rev.} {\bf D68}
(2003)123512, hep-th/0307288; hep-th/0412030;
E. Abdalla, S. Nojiri and S.D. Odintsov, hep-th/0409177.

\bibitem{stelle}
 K.S.~Stelle,
  Phys.\ Rev.\ D {\bf 16} (1977) 953.
\bibitem{barth}
  N.~H.~Barth and S.~M.~Christensen,
  Phys.\ Rev.\ D {\bf 28} (1983) 1876.

\bibitem{buch}
I.L. Buchbinder, S.D. Odintsov and I.L. Shapiro.
{ \em Effective Action in Quantum  Gravity} 
IOP Publishing, Bristol, 1992. 


\bibitem{f}
  G.~Cognola, E.~Elizalde, S.~Nojiri, S.~D.~Odintsov and S.~Zerbini,
  JCAP {\bf 0502} (2005) 010
  [arXiv:hep-th/0501096].


\bibitem{perry} G. W.Gibbson and M.J. Perry, 
{\sl Nucl. Phys. } {\bf B 146} (1978) 90.

\bibitem{duff80} S.M. Christensen and M.J. Duff, 
{\sl Nucl. Phys. } {\bf B 170} (1980) 480.


\bibitem{frad} E.S. Fradkin and A.A. Tseytlin, 
{\sl Nucl. Phys. } {\bf B 234} (1984) 472.

\bibitem{ds1}
S.D. Odintsov,{\sl Europhys.Lett.}{\bf 10} (1989) 287;
{\sl Theor.Math.Phys.}{\bf 82} (1990) 66;
T.R. Taylor and G. Veneziano, {\sl Nucl. Phys.} {\bf B345} (1990) 210;



\bibitem{eli94}
E. Elizalde, S.D. Odintsov, A. Romeo, A.A. Bytsenko and S. Zerbini.
{ \em Zeta regularization techniques with applications} 
World Scientific, 1994. 
\bibitem{byts96}
A.A. Bytsenko, G. Cognola, L. Vanzo  and S. Zerbini, 
{\sl Phys.Reports.} {\bf 269} (1996)1.


\bibitem{staro} A. Starobinsky, {\sl Phys.Lett.} {\bf B91} (1980) 99;
S.G. Mamaev and V.M. Mostepanenko, {\sl JETP} {\bf 51} (1980) 9;  B. Geyer, 
S. D. Odintsov and S. Zerbini, 
{\sl Phys. Letters} {\bf B 460} (1999) 58.


\bibitem{barrow} A.A.~Starobinsky, Sov.~Phys. - JETP Lett. {\bf 34}, 438 (1981);
J.D.~Barrow and A.C.~Ottewill, {\sl J. Phys. A: Math. Gen.} {\bf 16} (1983) 2757.


\bibitem{vanzo2}
  I.~Brevik, S.~Nojiri, S.~D.~Odintsov and L.~Vanzo,
  Phys.\ Rev.\ D {\bf 70}, 043520 (2004)
  [arXiv:hep-th/0401073]. 

\bibitem{Vanzo}
  L.~Vanzo,
    Phys.\ Rev.\ D {\bf 56}, 6475 (1997)
  [arXiv:gr-qc/9705004].

\bibitem{dick} R.~Dick, Gen.~Rel.~Grav. {\bf 36}, 217 (2004), gr-qc/0307052.

\bibitem{faraoni}
  V.~Faraoni,
    Phys.\ Rev.\ D {\bf 72}, 061501 (2005)
  [arXiv:gr-qc/9705004].


\bibitem{B} J.M. Bardeen, {\em Phys. Rev. D} {\bf 22},  
1882 (1980);
G.F.R. Ellis and M. Bruni, {\em Phys. Rev. D} {\bf 40}, 1804 
(1989); 
G.F.R. Ellis, J.-C. Hwang and M. Bruni, {\em Phys. Rev. D} 
{\bf 40},  1819 (1989);
G.F.R. Ellis, M. Bruni and J.-C. Hwang, {\em Phys. Rev. D} 
{\bf 42}, 1035 (1990).

\bibitem{odi2005}
  S.~Nojiri and S.~D.~Odintsov, arXiv:hep-th/0508049;
to appear in Phys.~Lett. B.
 


\end{thebibliography}
\end{document}